# Synthesis of titanate nanofibers co-sensitized with ZnS and $Bi_2S_3$ nanocrystallites and their application on pollutants removal

T.J. Entradas[1,a], J.C. Cabrita[1], B. Barrocas[1], M.R. Nunes[1], A.J. Silvestre[2], O.C. Monteiro[1,*]

[1] Centro de Química e Bioquímica, Faculdade de Ciências, Universidade de Lisboa, Campo Grande, 1749-016 Lisboa, Portugal.
[2] Department of Physics and CeFEMA, ISEL - Instituto Superior de Engenharia de Lisboa, Instituto Politécnico de Lisboa, Rua Conselheiro Emídio Navarro 1, 1959-007 Lisboa, Portugal.

**ABSTRACT**

The synthesis of nanocomposite materials combining titanate nanofibers (TNF) with nanocrystalline ZnS and $Bi_2S_3$ semiconductors is described in this work. The TNF were produced via hydrothermal synthesis and sensitized with the semiconductor nanoparticles, through a single-source precursor decomposition method. ZnS and $Bi_2S_3$ nanoparticles were successfully grown onto the TNF's surface and $Bi_2S_3$-ZnS/TNF nanocomposite materials with different layouts were obtained using either a layer-by-layer or a co-sensitization approach. The samples' photocatalytic performance was first evaluated through the production of the hydroxyl radical using terephthalic acid as probe molecule. All the tested samples show photocatalytic ability for the production of this oxidizing species. Afterwards, the samples were investigated for the removal of methylene blue. The nanocomposite materials with best adsorption ability for the organic dye were the ZnS/TNF and $Bi_2S_3$ZnS/TNF. The removal of the methylene blue was systematically studied, and the most promising results were obtained considering a sequential combination of an adsorption-photocatalytic degradation process using the $Bi_2S_3$ZnS/TNF powder as a highly adsorbent and photocatalyst material.

**Keywords:** $Bi_2S_3$-ZnS/TNF nanocomposites; photosensitization; hydroxyl radical; pollutants removal.

---

[a]Present address: Istituto per lo Studio delle Macromolecole, Via Edoardo Bassini, 15, 20133 Milano, Italy
[*]Corresponding author: Phone:+351 217500865; Fax:+351 217500088; E-mail: ocmonteiro@fc.ul.pt

# 1. Introduction

In the past decades there has been a growing interest in photo-assisted catalysis using semiconductors as an advanced oxidation process for the elimination of many organic pollutants in wastewater systems [1]. However, the complete mineralization of a large variety of organic compounds with formation of $CO_2$, $H_2O$ and mineral acids, is not always easy to be achieved using only a photocatalysis process. To overcome this problem, the integration of different methodologies to achieve the complete removal of industrial pollutants from wastewater systems is nowadays the subject of great interest [2-5].

Photocatalytic methods are generally based on the generation of oxidizing species, namely OH• radicals, which interact with organic pollutants, leading to their progressive degradation and subsequently mineralization. During the irradiation of a photocatalyst, the photo-generation of electron-hole pairs is possible once the photons' energy is higher than the bandgap energy of the irradiated material. These charge carriers can participate in redox reactions at the semiconductor surface, forming reactive radicals that are able to degrade the pollutants. In order to improve the performance of the photodegradation processes, heterogeneous photocatalysis can be combined with other pollutant removal technologies [6], allowing an increase of the global process efficiency, by decreasing the reaction time and/or costs. For instance, by combining heterogeneous photocatalysis with ultrasonic irradiation [7], photo-Fenton reaction [8], or electrochemical treatment [9], the photocatalytic mechanisms are directly affected. On the other hand, by associating photocatalysis with biological treatments [4,5], physical adsorption [10], or membrane reactor [11], the overall degradation process efficiency can be improved, although no modification in the photocatalytic mechanism is expected to occur.

In an adsorption–photocatalysis combined process, an appropriated adsorbent, *e.g.* silica, alumina, zeolites or activated carbon, will selectively adsorb the pollutant present in the wastewater. Once the adsorbent is saturated, losing its ability to remove the organic contaminant, it will be separated from the cleaned/treated water. Subsequently it can be added to a photocatalyst in an aqueous medium to produce a suspension. By conveniently irradiating this suspension, the pollutant will be photocatalytically oxidized and the adsorbent material can be regenerated and reused for further adsorption cycles. This process can be considerably simplified by using a material that simultaneously combines good adsorption and photocatalytic properties. In such case, after total pollutant adsorption (adsorption step) the irradiation of the adsorbent will result in the pollutant degradation and in the adsorbent/catalyst recovery (photocatalytic step). This stepwise process is a good strategy to achieve the remediation of wastewater systems without irradiation as a prerequisite, the recovering of the adsorbent/catalyst material taking subsequently place in a much more practical manner. One possible approach to develop such materials consists in combining a photocatalyst with a highly adsorbent material, thus producing a catalyst–adsorbent nanocomposite material where adsorption and oxidation of organic compounds can occur.

Among the prospective materials to be used in this context, titanate nanotubes and titanate nanofibers (TNF) can be very interesting because they combine properties of conventional $TiO_2$ nanoparticles (*e.g.* photocatalytic activity) with the properties of layered titanates (*e.g.* ion-exchange facility). Consequently, they have the potential to be used in a wide variety of applications, including photocatalysis [12-14] dye-sensitized solar cells [15,16], transparent optical devices [17-19], and gas or humidity sensors [20,21]. Particularly important could be their use in photocatalytic applications since they have a similar structure to titanium dioxide but a higher specific surface area due to their open mesoporous morphology. Nonetheless, they present also drawbacks: their high charge recombination rate and wide band gap (*ca.* 3.3 eV), which limits the electron-hole photo-generation under visible irradiation. Therefore, the synthesis of TNF-*based* nanocomposite materials with either a broader range of light absorption and/or a lower charge recombination rate would be a significant achievement towards the development of successful photocatalyst materials. For instance, the combination of TNF with active materials in the visible range of the electromagnetic spectrum such as nanocrystalline $Bi_2S_3$, could extend the optical absorption of the nanocomposite material. On the other hand, a lower recombination rate of the photo-generated carriers, may be achieved by combining TNF with ZnS, due to interesting optical properties of this direct semiconductor [22-24].

This work reports on the synthesis of TNF-*based* nanocomposite materials combining titanate nanofibers with nanocrystalline $Bi_2S_3$ and/or ZnS semiconductors, and on their potential use for pollutant removal. Titanate nanofibers were prepared using a hydrothermal treatment of an amorphous precursor [13]. The synthesis of ZnS-$Bi_2S_3$/TNF nanocomposites was performed using organometallic compounds, metal dithiocarbamato complexes, as single-sources for the metal sulphides grown on the TNF's surface [2,25].

The photocatalytic activity of the prepared nanocomposite samples was investigated using the methylene blue (MB) degradation process. As it is well-known, MB is a bright coloured cationic thiazine dye that is extensively used in textile industries, beyond its use in biology, medicine and veterinary medicine, where is commonly used in *in-vivo* diagnosis. The strong affinity of MB to the nanocomposite samples' surface results on a rapid and efficient method for the decolourization of aqueous solutions without any special instrumental equipment request, *i.e.* optical lamp system. Subsequently, the irradiation of the mixture leads to the degradation of the dye, ensuing simultaneously a total recovery of the adsorbent/photocatalyst nanocomposite material.

## 2. Experimental

### 2.1. Materials

All reagents were of analytical grade and were used as received without further purification. The solutions were prepared with bi-distilled water.

*2.1.1. TNF synthesis*

The TNF samples were prepared via the hydrothermal method described elsewhere [13,14]. A titanium trichloride solution (10 wt.% in 20-30 wt.% HCl) diluted in a ratio of 1:2 in standard HCl solution (37 %) was used as the titanium source. To this solution a 4 M ammonia aqueous solution was added drop-wise under vigorous stirring, until complete precipitation of a white solid. The resulting suspension (precursor) was kept overnight at room temperature and then filtered and vigorously rinsed with deionised water in order to remove the remaining ammonia and chloride ions.

The TNF synthesis was performed in an autoclave system using ~6 g of the precursor in *ca.* 60 ml of NaOH 10 M aqueous solution. The samples were prepared at 200 ºC using an autoclave dwell time of 12 h. After cooling, the suspension was filtrated and the white solid was washed until pH = 7 has been reached in the filtrate solution.

*2.1.2 TNF sensitization with nanocrystalline ZnS and $Bi_2S_3$*

The ZnS/TNF nanocomposite samples were prepared based on a procedure previously reported [26]. Accordingly, the TNF were sensitized with ZnS by adding 2.5 mL of ethylenediamine to 50 mL of a acetone suspension containing 0.125 mmol of $\{Zn[S_2CN(C_2H_5)_2]_2\}$ and 0.250 g of TNFs. The $Bi_2S_3$/TNF samples were prepared by using $\{Bi_2[S_2CN(C_2H_5)_2]_3\}$ as bismuth sulphide precursor. The suspensions were stirred at reflux temperature, during 4 hours. After cooling to room temperature, the ZnS/TNF (crème solid) and $Bi_2S_3$/TNF (dark brown solid) were isolated by centrifugation, washed with acetone and dried at room temperature in a desiccator over silica gel.

The $Bi_2S_3$(ZnS/TNF), ZnS($Bi_2S_3$/TNF) and $Bi_2S_3$ZnS/TNF nanocomposite materials were prepared by adjusting the experimental procedure as follows: *i)* the $Bi_2S_3(ZnS/TNF)$ samples were produced after preparing the ZnS/TNF powders and then using them as matrix for the $Bi_2S_3$ sensitization; *ii)* the $ZnS(Bi_2S_3/TNF)$ samples were obtained in a similar mode by sensitizing $Bi_2S_3$/TNF powders with ZnS; *iii)* the $Bi_2S_3ZnS$/TNF samples were produced using simultaneously both metal sulphide precursors during a single step sensitization process.

*2.1.3. ZnS and $Bi_2S_3$ precursors*

Zinc diethyldithiocarbamate, $\{Zn[S_2CN(C_2H_5)_2]_2\}$, was prepared as previously reported [27]: 20 mmol of ethylenediamine ($C_2H_8N_2$) and 13.5 mmol of carbon disulphide ($CS_2$) were added to a suspension containing 10 mmol of $ZnCl_2$ in water (50 mL). The mixture was stirred over 2 hours and a white solid was obtained. The solid was crystallized in hot chloroform or dichloromethane.

Bismuth diethyldithiocarbamate, $\{Bi_2[S_2CN(C_2H_5)_2]_3\}$, was prepared as reported [28]: 40 mmol of $C_2H_8N_2$ and 50 mmol of $CS_2$ were added to a suspension containing 6 mmol of bismuth(III) oxide ($Bi_2O_3$) in methanol (20 mL). The mixture was stirred over 24 h and a yellow solid was obtained. The solid was recrystallized in hot chloroform/methanol (3:1).

After recrystallization, both metal complexes were identified using FTIR.

*2.1.4. ZnS and $Bi_2S_3$ crystalline nanoparticles synthesis*

For comparative proposes, samples of crystalline ZnS, $Bi_2S_3$ and $Bi_2S_3ZnS$ nanoparticles were prepared as described in Section 2.1.2, in the absence of titanate nanofibers. For the sake of clarity, these samples were labelled nanoZnS, nano$Bi_2S_3$ and nano($Bi_2S_3ZnS$), respectively.

**2.2 Photodegradation experiments**

All photodegradation experiments were conducted using a 250 ml refrigerated photoreactor [29]. A 450W Hanovia medium-pressure mercury-vapour lamp was used as radiation source, the total irradiated energy being 40-48% in the ultraviolet range and 40-43% in the visible region of the electromagnetic spectrum. Prior to irradiation, the suspensions were stirred in darkness conditions during 60 min to ensure the adsorption equilibrium. During irradiation, suspensions were sampled at regular intervals, centrifuged and analysed by UV-vis or fluorescence spectroscopy.

The evaluation of the hydroxyl radical production was carried out using terephthalic acid (TA) as probe molecule. Suspensions prepared using 150 mL of a TA solution (10 mM, in 0.001 M NaOH) and 10 mg of each powder were used during the photocatalytic experiments. For the photocatalytic experiments, methylene blue suspensions were prepared by adding 25 mg of the powder to 125 ml of a dye aqueous solution (6 ppm).

## 2.3. Characterization

X-ray powder diffraction was performed using a Philips X-ray diffractometer (PW 1730) with automatic data acquisition (APD Philips v3.6B), using Cu Kα radiation ($\lambda$ = 0.15406 nm) and working at 40 kV/30 mA. The diffraction patterns were collected in the $2\theta$ range of 20º - 60º with a 0.02º step size and an acquisition time of 200 s/step. Transmission electron microscopy (TEM), high resolution transmission electron microscopy (HRTEM) and selected area electron diffraction (SAED) were carried out on a JEOL 200CX microscope operating at 200 kV. The elemental composition of the samples were analysed by energy dispersive X-ray spectroscopy (EDS). UV–VIS absorption spectra of the solutions were recorded using a UV–VIS absorption spectrophotometer (Shimadzu, UV-2600PC). The powder diffuse reflectance spectra (DRS) were recorded in the wavelength range of 220-1400 nm using the same equipment with an ISR 2600plus integration sphere. Fluorescence spectra of 2-hydroxyterephthalic acid were measured on a Spex Flourolog 3-22/Tau 3 fluorescence spectrophotometer. The excitation was performed at 315 nm and the emission at 425 nm.

## 3. Results and discussion

### 3.1. ZnS/TNF and Bi$_2$S$_3$/TNF samples characterization

*3.1.1. Crystalline structure and morphology*

Figure 1 shows the diffractograms of the prepared TNF-*based* nanocomposite samples as well as that of the *non*-sensitized TNF, nanoZnS and nanoBi$_2$S$_3$ samples, for comparative proposes. As can be seen, the XRD patterns of both ZnS/TNF and Bi$_2$S$_3$/TNF show the typical $2\theta \sim 10º$ peak, which is related with the interlayer distance of the TNF [13]. Therefore, the sensitization process preserves the TNF crystalline structure. Concerning the ZnS/TNF sample, the XRD pattern shows a wide peak at $2\theta \sim 28°$ resulting from the overlap of the (100) (002) and (101) ZnS diffraction planes centred at $2\theta = 26.914°$, 28.494° and 30.538°, respectively (*wurtzite* crystal structure, JCPDS nº10-0443). This result is in agreement with the presence of extremely small ZnS crystallites over the TNF's surface. Similarly, the XRD pattern of the Bi$_2$S$_3$/TNF sample is in accordance with the presence of nanocrystalline Bi$_2$S$_3$ particles (*bismuthinite* crystal structure, JCPDS file 17-0320) over the TNF particles' surface. Moreover, these data agree with previous related works [2,28].

The morphology of the samples was studied by transmission electronic microscopy. Prior to sensitization, the TNF sample is morphologically homogeneous and comprises thin elongated nanoparticles with a high length/diameter aspect ratio (not shown) [14]. Figure 2a shows a TEM image of the ZnS/TNF sample. As depicted, a homogeneous thin layer of a second material is recovering the entire TNF surface. A close inspection of the image (Figure 2b) shows that this shell-layer is constituted by extremely small particles, whose morphology is identical to that obtained for the nanoZnS sample (not shown). The presence of ZnS particles was furthermore confirmed by EDS analysis, showing the existence of Zn and S elements in a semi-quantitative 1:1 ratio (not shown). The crystallinity of the small ZnS particles, forming the thin layer over the TNF surface was confirmed by SAED analysis (Figure 2c).

Likewise, the morphology of the $Bi_2S_3$-*based* samples were also analysed by TEM. Figure 3a shows a TEM micrograph of a $Bi_2S_3$/TNF sample. A detail of this image is shown in Figure 3b. As can be seen, lamellar-*like* particles have grown onto the TNF's surface after decoration with $Bi_2S_3$, and no evidence of segregation or other phase's formation was observed. This type of structure is expected since $Bi_2S_3$ is a semiconductor that can easily present preferential growth according to a particular plane, often resulting in materials exhibiting lamellar morphology. An identical morphology was observed for the nano$Bi_2S_3$ powder, which presents a sheet-like network morphology leading, by energetic reasons, to the formation of spherical agglomerates [30]. The SAED analysis of one single decorated composite particle indicates the presence of two different crystalline phases that are consistent with the existence of crystalline $Bi_2S_3$ nanosheets covering the TNF surface (Figure 3c). The EDS analysis performed over a $Bi_2S_3$/TNF sample reveals the presence of Bi and S elements, beyond Ti and Na from TNF (not shown).

### 3.1.2. Optical characterization

In order to study the effect of the nanocrystalline semiconductor layer in the optical properties of the TNF, the diffuse reflectance (*R*) spectra of the prepared samples were recorded. *R* is related with the absorption Kubelka-Munk function, $F_{KM}$, by the relation $F_{KM}(R) = (1-R)^2/2R$, which is proportional to the absorption coefficient [31]. Figure 4a shows the absorption spectra of the ZnS-*based* samples. As expected, the TNF, nanoZnS and ZnS/TNF samples start to absorb radiation only in the near UV range. As can be observed, the absorption band edge of the ZnS/TNF sample is blue shifted in comparison to that of nanoZnS sample. This blue shift is related with the relative ZnS nanoparticles size; meaning that the ZnS nanocrystallites are slight smaller when grown at the TNF's surface. Note, however, that the ZnS/TNF sample has also a slightly absorption increase on the visible radiation range. The samples' band gap energies ($E_g$) were estimated through the Kubelka-Munk function following the procedures described elsewhere [13]. Bandgap energies of 3.82, 3.61 and 3.76 eV were estimated for the TNF, nanoZnS and ZnS/TNF samples respectively.

The absorption spectra of the $Bi_2S_3$-*based* samples are shown in Figure 4b. In the spectrum of sample $Bi_2S_3$/TNF two band edges can be seen, one near 335 nm and the other starting at 1050 nm. The former is attributed to the TNF absorption and the latter is related to the $Bi_2S_3$ particles. This result agrees with the existence of a layered nanocomposite material in which the second semiconductor phase covers completely the TNF surface. However, the $Bi_2S_3$ is present as a very thin layer, still allowing some TNF' light absorption. The shoulder observed in the $Bi_2S_3$/TNF spectrum in the range 650-850 nm may be related with new electronic states within the TNF forbidden band [32]. It is interesting to note that for $\lambda > 850$ nm, the optical absorption behaviour attributed to the $Bi_2S_3$ is similar for both nano$Bi_2S_3$ and $Bi_2S_3$/TNF samples, suggesting that this nanocrystalline semiconductor growth process is independent on the TNF presence/absence. $E_g$ values of 1.24 eV and 1.28 eV were estimated for samples nano$Bi_2S_3$ and $Bi_2S_3$/TNF, respectively.

### 3.2. $Bi_2S_3ZnS$/TNF, $ZnS(Bi_2S_3$/TNF) and $Bi_2S_3(ZnS$/TNF) samples characterization

*3.2.1. Crystalline structure and morphology*

Figure 5 presents the diffractograms of the samples $Bi_2S_3ZnS$/TNF (TNF sensitized simultaneously with ZnS and $Bi_2S_3$), $ZnS(Bi_2S_3$/TNF) ($Bi_2S_3$/TNF sensitized with ZnS) and $Bi_2S_3(ZnS$/TNF) (ZnS/TNF sensitized with $Bi_2S_3$). As can be seen, all the three samples show similar XRD patterns from which the presence of nanocrystalline ZnS and $Bi_2S_3$ along with the TNF can be inferred, taking into account the XRD analysis of the ZnS/TNF and $Bi_2S_3$/TNF samples previously discussed in section 3.1.1.

Figure 6 shows the TEM micrographs of samples $Bi_2S_3ZnS$/TNF, $ZnS(Bi_2S_3$/TNF) and $Bi_2S_3(ZnS$/TNF). As can be seen, the three nanocomposite materials present similar morphology. The existence of a second-layer covering the TNF's surface is clearly seen in all samples; this layer somehow resembles to that of $Bi_2S_3$/TNF, suggesting the presence of nanocrystalline $Bi_2S_3$ over the TNF's surface in all the three co-sensitized samples. The undoubting presence of small spheroidal ZnS nanocrystallites in the three nanocomposite samples was confirmed through the higher magnified TEM images also shown in Figure 5.

*3.2.2. Optical characterization*

Figure 7 shows the absorption spectra of the three co-sensitized samples. As shown, all samples absorb in the entire visible range. This is due to the presence of nanocrystalline $Bi_2S_3$, which is a narrow bandgap semiconductor ($E_g = 1.08$ eV). Moreover, since the absorption of the three samples is dominated by the nanocrystalline $Bi_2S_3$, the bandgap of the co-sensitized samples, (1.30 eV, 1.35 eV and 1.23 eV for $Bi_2S_3(ZnS/TNF)$, $ZnS(Bi_2S_3/TNF)$ and $Bi_2S_3ZnS/TNF$ samples, respectively) are close to the bandgap inferred for the nano$Bi_2S_3$ sample (1.34 eV).

## 3.3. Hydroxyl radical photocatalytic production evaluation

In order to investigate the prepared nanocomposite materials for future photocatalytic applications, the evaluation of the catalytic production of hydroxyl radical formation (·OH) was carried out. The quantification of the hydroxyl radical production was performed using terephthalic acid (TA) as probe molecule. Terephthalic acid is a well-known ·OH scavenger, which does not react with other radicals, such as $O_2^-$, $HO_2$ and $H_2O_2$. The ·OH radical can converts terephthalic acid in to the fluorescent 2- hydroxytherephthalic acid (HTA) through the reaction

$$C_6H_4(COOH)_2 + \cdot OH \rightarrow C_6H_4(COOH)_2OH \tag{1}$$

First, and for comparative proposes, the nanoZnS, nano$Bi_2S_3$ and nano($Bi_2S_3ZnS$) samples were tested for the TA photocatalytic degradation, during 25 minutes of irradiation. The results obtained concerning the HTA formation are present in Figure 8a. As can be observed, the highest amount of HTA produced was achieved using the nanoZnS sample as catalyst. A lower performance was attained with the nano($Bi_2S_3ZnS$) sample and no catalytic activity for reaction (1) was observed using the nano$Bi_2S_3$ sample. Nevertheless, the use of nanocrystalline ZnS as photocatalyst for future real world applications has two main drawbacks: the reduced amount of powder produced per batch and the difficult solid-solution separation after photocatalysis. These drawbacks make the use of nanoZnS as a photocatalyst an expensive and time consuming process that is very difficult to scale-up to an industrial level.

Bearing in mind these facts, the ZnS/TNF was also evaluated for the TA photodegradation. After 25 minutes of irradiation, the HTA formation was similar for the nanoZnS and ZnS/TNF samples (Figure 8b). It should be notice that the total amount of nanocrystalline photoactive ZnS used in the nanocomposite ZnS/TNF sample is much lower than that present in the nanoZnS sample. Moreover, the recovery of the composite solid after photocatalysis is easier since the particles are larger, making this material much more attractive for industrial applications then the nanoZnS *per se*.

Figure 8c shows the hydroxyl radical production in the presence of the co-photosensitized samples. After 25 min of irradiation, the highest amount of HTA was achieved by using the $Bi_2S_3ZnS/TNF$ powder as photocatalyst. Moreover, the production of HTA via reaction (1) continuously increases over time in the presence of this nanocomposite material. When using the $ZnS(Bi_2S_3/TNF)$ sample as photocatalyst, the production of HTA increases rapidly up to 10 min, reaching a quasi-equilibrium state for longer irradiation times which corresponds to a production of HTA lower than that achieved in the presence of $Bi_2S_3ZnS/TNF$. The worst photocatalytic performance was observed for the $Bi_2S_3(ZnS/TNF)$ powder, the amount of HTA produced in the presence of this sample being similar to that produced during the photolysis. This behaviour may probably be attributed to the ZnS' layer inner position in this nanocomposite material.

### 3.4. Methylene blue photocatalytic degradation

Taking into account the above results, the samples TNF, ZnS/TNF and $Bi_2S_3ZnS/TNF$ were selected for testing MB removal. Samples were first evaluated for the dye adsorption in dark conditions. The TNF sample didn't present any ability to adsorb MB while ZnS/TNF sample adsorbs 17 % of the dye. On the other hand, when using $Bi_2S_3ZnS/TNF$ sample no dye was observed in the MB solution after 15 minutes in dark conditions, with the dye becoming completely adsorbed by this nanocomposite material in a short period of time. Note that a similar result was observed when using the $Bi_2S_3/TNF$ sample. These results show that the dye adsorption takes place almost in the nanocrystalline $Bi_2S_3$ since no MB adsorption occur using pristine TNF.

The photocatalytic activities of the ZnS/TNF and $Bi_2S_3ZnS/TNF$ nanocomposite materials were monitored by measuring the maximum absorbance intensity of the MB's chromophoric peak ($\lambda$ = 664 nm) of a 6 ppm MB aqueous solution, as a function of the irradiation time in the presence of the photocatalyst materials. Prior to irradiation, the suspensions were stirred in darkness conditions during 60 min in order to ensure the adsorption equilibrium. The best photocatalytic performance was obtained for the ZnS/TNF material. Indeed, after 60 min of irradiation almost no MB remains in solution, as shown in the absorption spectra of Figure 9, where a decrease of the MB's chromophoric absorption peak down to nearly zero can be seen as the irradiation time increases up to 60 min. However, after this period the solid had a blue colour indicating that some dye stay adsorbed in the catalyst surface and, consequently, an additional irradiation time will be required to conclude the photodegradation process and to complete recover the photocatalyst.

The MB aqueous solution containing $Bi_2S_3ZnS/TNF$ was also irradiated. In this case, as mentioned above, all the dye was adsorbed by the photocatalyst material during the period in which the suspension was kept in the dark. Figures 10a and 10b display the absorption spectra of this suspension up to 180 min after starting irradiation. Figure 10c shows the relative intensity ($I/I_0$) of the MB absorption chromophoric peak at $\lambda = 664$ nm of the suspension, plotted as a function of the irradiation time. $I_0$ stands for the band intensity of the 6 ppm MB reference solution. As can be seen, the MB concentration in the solution continually increases during the first 45 min of irradiation. This result indicates that irradiation disrupts the adsorption equilibrium reached prior to turning on the light, part of the previously adsorbed MB being gradually released into solution alongside with the MB photodegradation reaction. After this period, the rate at which the MB photodegradation is taking place is higher than that at which MB desorption occurs, and consequently the dye concentration starts to decrease. The complete decomposition of MB and total recovery of the $Bi_2S_3ZnS/TNF$ adsorbent/catalyst is reached for an irradiation time of 180 min. These are very promising results since they indicate that the $Bi_2S_3ZnS/TNF$ nanocomposite can be used to remove large amounts of MB from aquatic media without the prerequisite of an *in-situ* light source. Indeed, after achieving the maximum adsorption capacity, the nanocomposite can be removed and subsequently recovered in an adequate *ex-situ* facility by using a photo-assisted dye degradation process. These two independent processes can be very attractive from an economical point of view for industrial applications.

## 4. Conclusions

New nanocomposite materials combining crystalline TNF with nanocrystalline ZnS and $Bi_2S_3$ semiconductors were prepared. ZnS and $Bi_2S_3$ nanoparticles were grown onto the TNF's surface by using either a layer-by-layer or a co-sensitization approach through a single-source precursor methodology. $Bi_2S_3$-ZnS/TNF nanocomposites with different layouts were obtained. The prepared samples were morphological, structural and optical characterized. The photocatalytic performance of the different samples was first evaluated throughout the production of the ·OH radical using terephthalic acid as probe molecule. All prepared samples show photocatalytic ability for the production of that oxidizing species. After, the adsorption and photocatalytic activity of the ZnS/TNF and $Bi_2S_3$ZnS/TNF samples were investigated for the methylene blue degradation process. ZnS/TNF reveals adsorption ability for the MB and is photocatalytic active for its degradation. However, from the industrial point of view, the $Bi_2S_3$ZnS/TNF nanocomposite seems to be the most interesting material to be used in heterogeneous photocatalysis. It has a strong ability to adsorb the organic dye and, therefore, to effectively remove it from aqueous solutions. The subsequent dye photodegradation and recovery of the photocatalyst material can be handled in a suitable *ex-situ* facility.

## Acknowledgements

This work was supported by Fundação para a Ciência e Tecnologia (FCT) under project No. PTDC/CTM-NAN/113021/2009 and PEst-OE/QUI/UI0612/2014.

# Figure captions

**Figure 1.** XRD patterns of the nanoZnS, nanoBi$_2$S$_3$, Bi$_2$S$_3$/TNF and ZnS/TNF samples. Inset: XRD pattern of the TNF powder. Vertical lines are the ZnS *wurtzite* crystal structure, JCPDS file n.º 10-0443 and Bi$_2$S$_3$ *bismuthinite* crystal structure, JCPDS file n.º 17-0320.

**Figure 2.** TEM images of a) ZnS/TNF particles, b) detail of a ZnS/TNF surface, and c) SAED pattern of a single ZnS/TNF particle.

**Figure 3.** TEM images of a) Bi$_2$S$_3$/TNF particle, b) detail of a Bi$_2$S$_3$/TNF surface, and c) SAED pattern of a single Bi$_2$S$_3$/TNF particle.

**Figure 4.** Absorption spectra of a) ZnS/TNF, TNF and nanoZnS samples, and b) Bi$_2$S$_3$/TNF, TNF and nanoBi$_2$S$_3$ samples.

**Figure 5.** XRD patterns of the Bi$_2$S$_3$ZnS/TNF, ZnS(Bi$_2$S$_3$/TNF), and Bi$_2$S$_3$(ZnS/TNF) co-sensitized samples.

**Figure 6.** TEM images of the Bi$_2$S$_3$ZnS/TNF, ZnS(Bi$_2$S$_3$/TNF), and Bi$_2$S$_3$(ZnS/TNF) co-sensitized samples.

**Figure 7.** Absorption spectra of the Bi$_2$S$_3$ZnS/TNF, ZnS(Bi$_2$S$_3$/TNF), and Bi$_2$S$_3$(ZnS/TNF) co-sensitized samples.

**Figure 8.** Hydroxyl radical formation resulting from the photodegradation of a terephthalic acid solution, using different photocatalysts.

**Figure 9.** Absorption spectra of a 6 ppm MB solution during irradiation using ZnS/TNF as photocatalyst.

**Figure 10.** a) Absorption spectra of a 6 ppm MB solution before and during irradiation using Bi$_2$S$_3$ZnS/TNF as adsorbent and photocatalyst, b) detail of the spectra shown in a), and c) relative intensity of the MB chromophoric absorption band at λ = 664 nm as a function of the irradiation time.

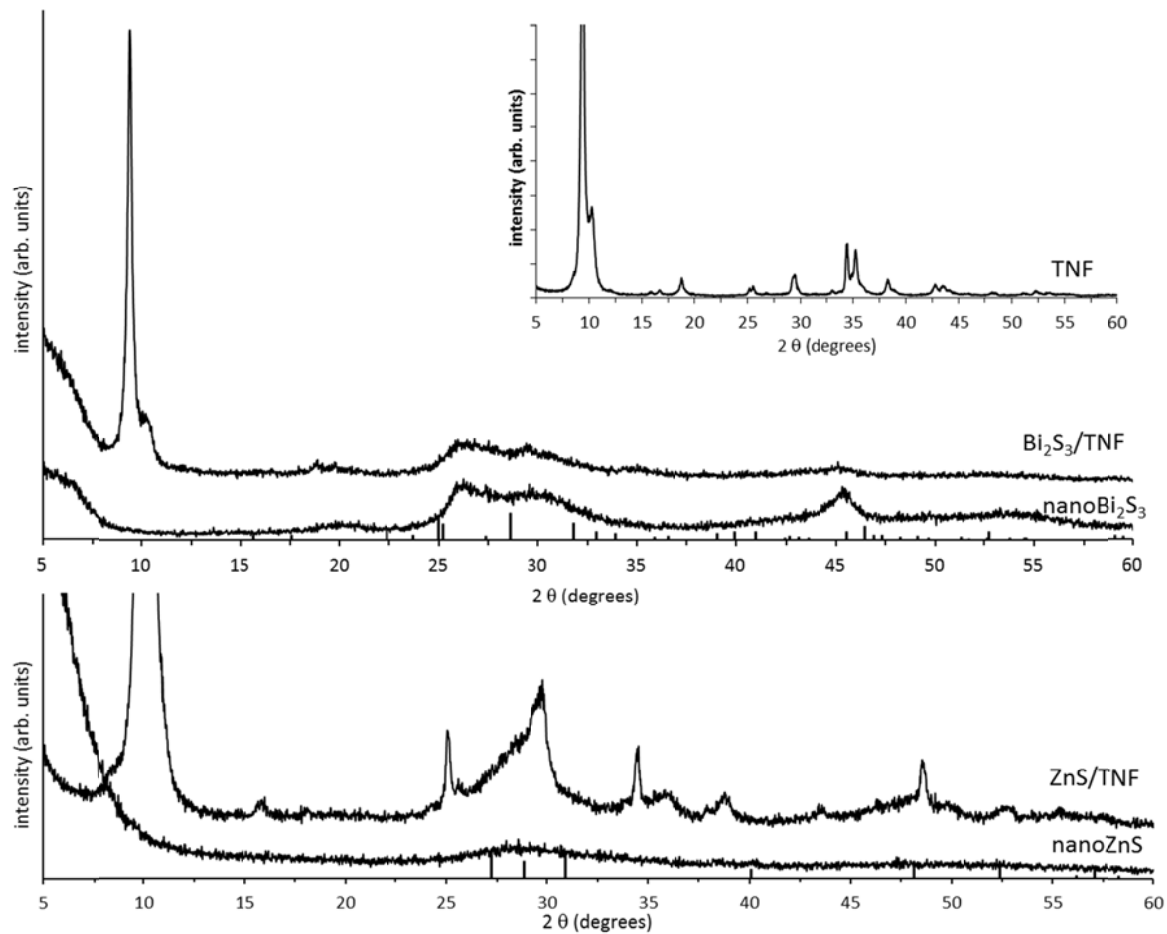

**Figure 1** – XRD patterns of the nanoZnS, nanoBi$_2$S$_3$, Bi$_2$S$_3$/TNF and ZnS/TNF samples. Inset: XRD pattern of the TNF powder. (vertical lines are the ZnS *wurtzīte* crystal structure, JCPDS file n°10-0443 and Bi$_2$S$_3$ *bismuthinite* crystal structure, JCPDS file 17-0320).

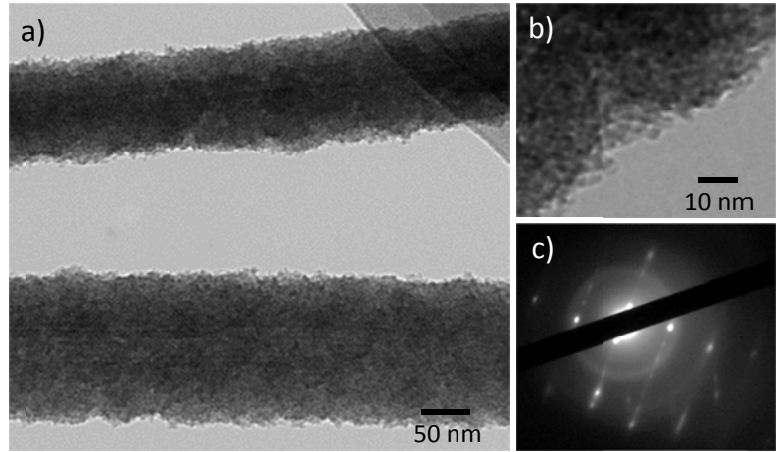

**Figure 2.** TEM images of a) ZnS/TNF particles, b) detail of a ZnS/TNF surface, and c) SAED pattern of a single ZnS/TNF particle.

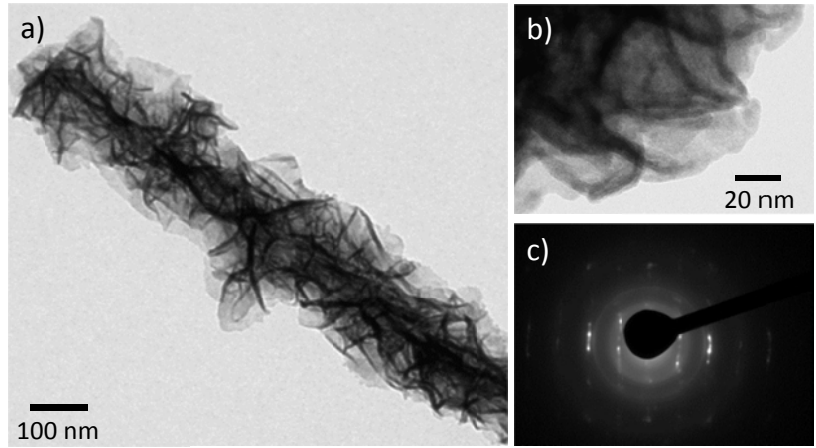

**Figure 3.** TEM images of a) $Bi_2S_3$/TNF particle, b) detail of a $Bi_2S_3$/TNF surface, and c) SAED pattern of a single $Bi_2S_3$/TNF particle.

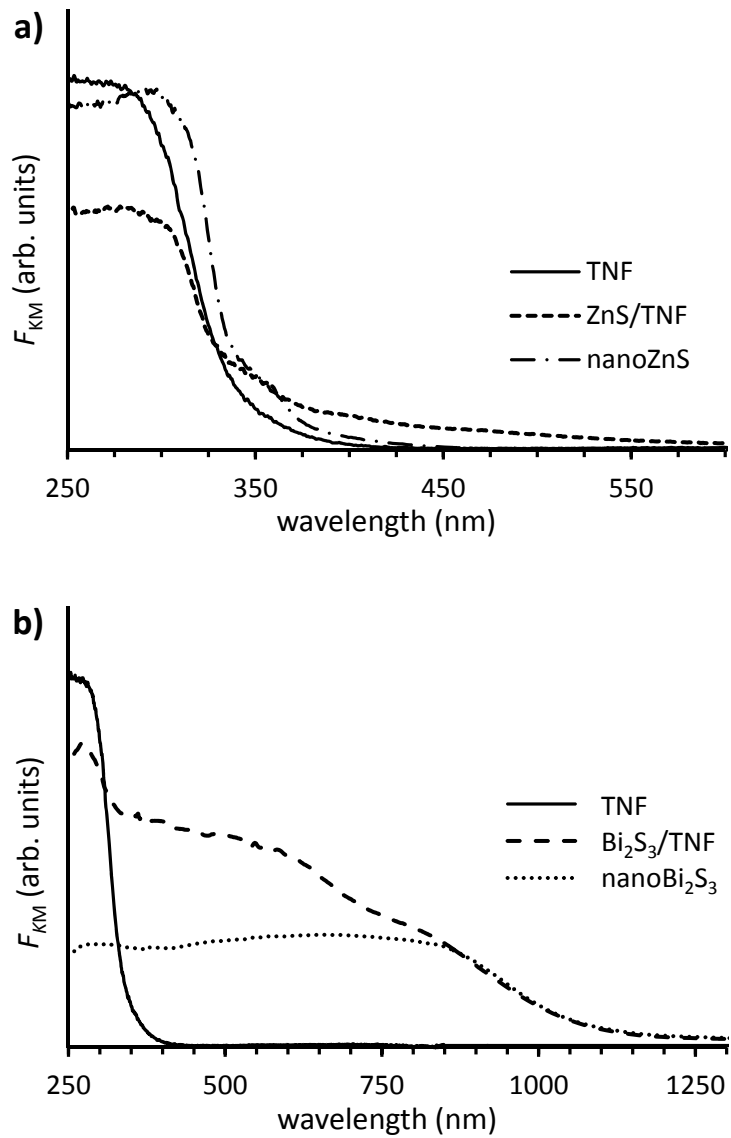

**Figure 4.** Absorption spectra of a) ZnS/TNF, TNF and nanoZnS samples, and b) $Bi_2S_3$/TNF, TNF and nano$Bi_2S_3$ samples.

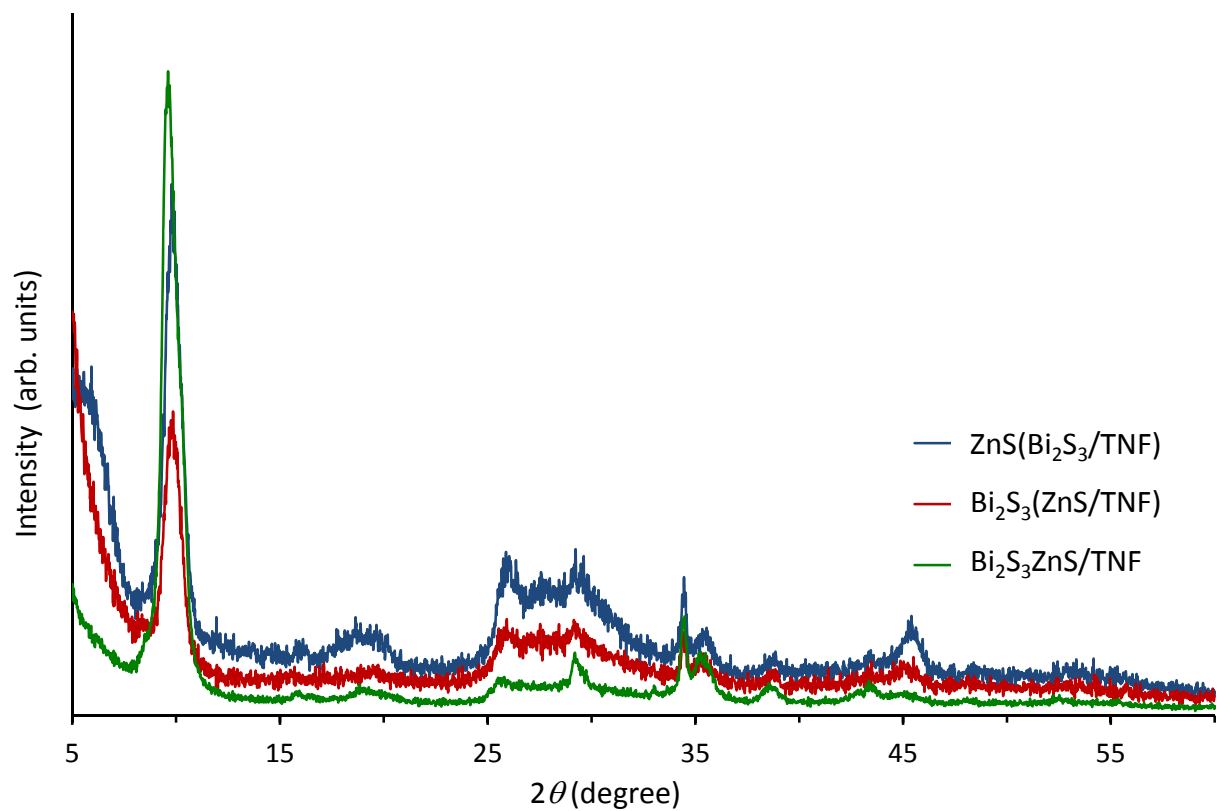

**Figure 5.** XRD patterns of the Bi$_2$S$_3$ZnS/TNF, ZnS(Bi$_2$S$_3$/TNF) and Bi$_2$S$_3$(ZnS/TNF) co-sensitized samples.

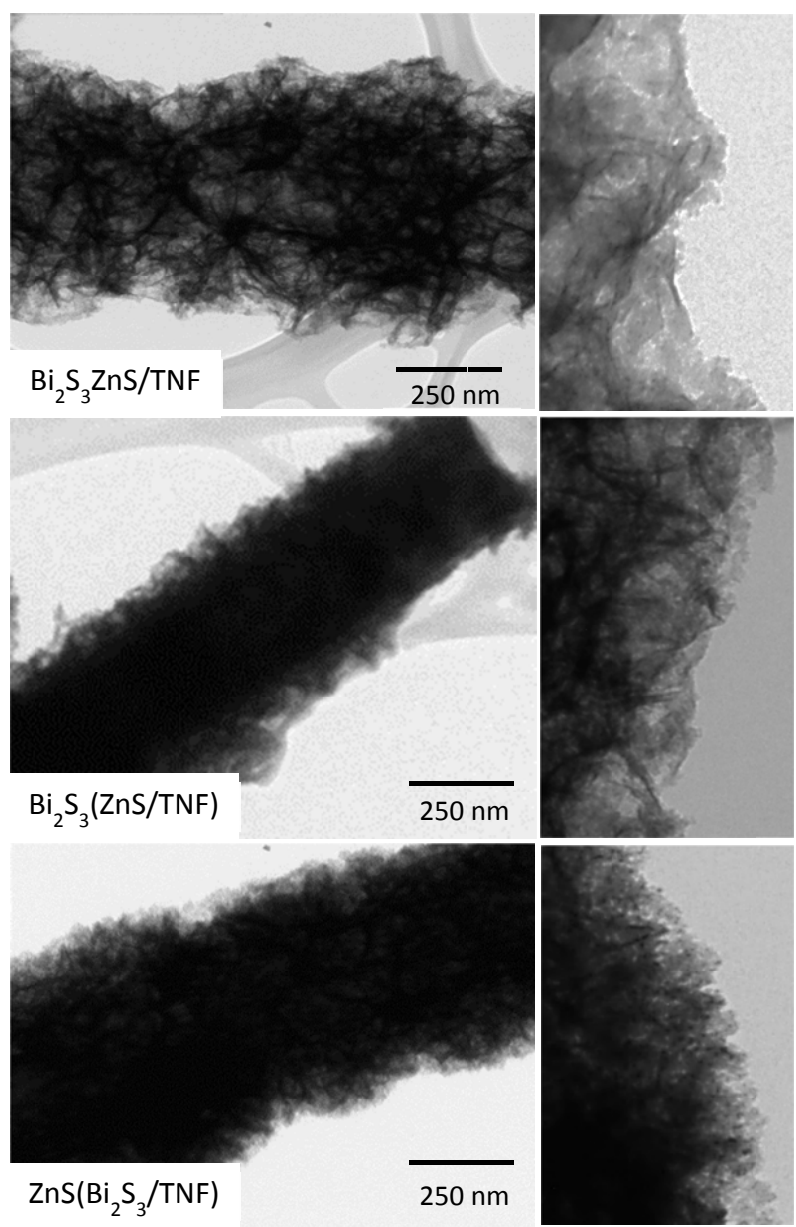

**Figure 6.** TEM images of the Bi$_2$S$_3$ZnS/TNF, ZnS(Bi$_2$S$_3$/TNF), and Bi$_2$S$_3$(ZnS/TNF) co-sensitized samples.

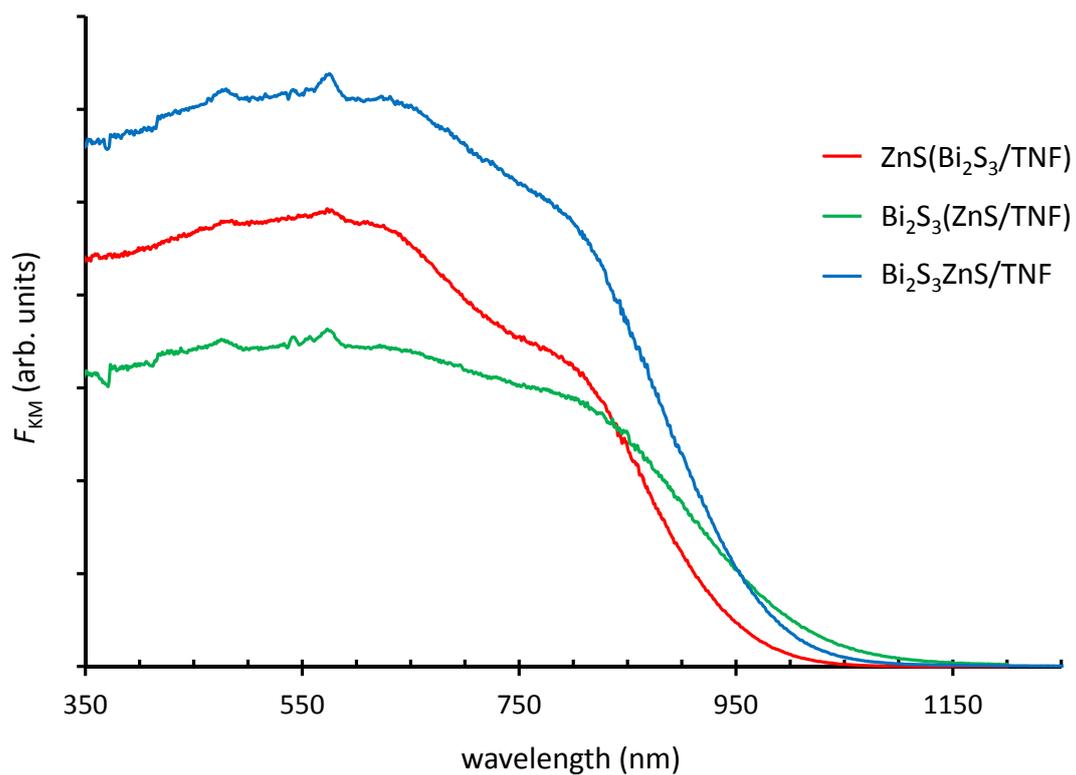

**Figure 7.** Absorption spectra of the Bi$_2$S$_3$ZnS/TNF, ZnS(Bi$_2$S$_3$/TNF), and Bi$_2$S$_3$(ZnS/TNF) co-sensitized samples.

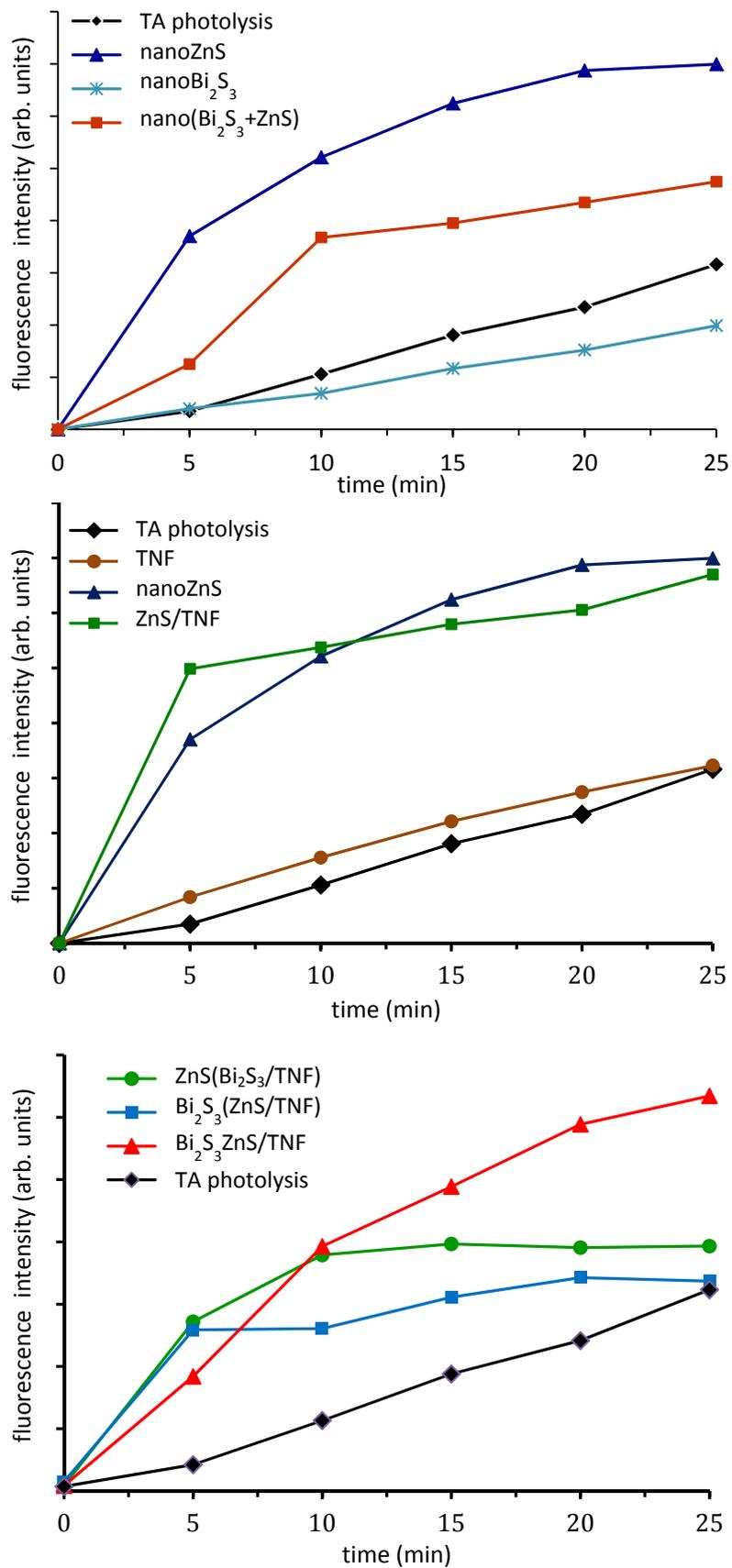

**Figure 8.** Hydroxyl radical formation resulting from the photodegradation of a terephthalic acid solution, using different photocatalysts.

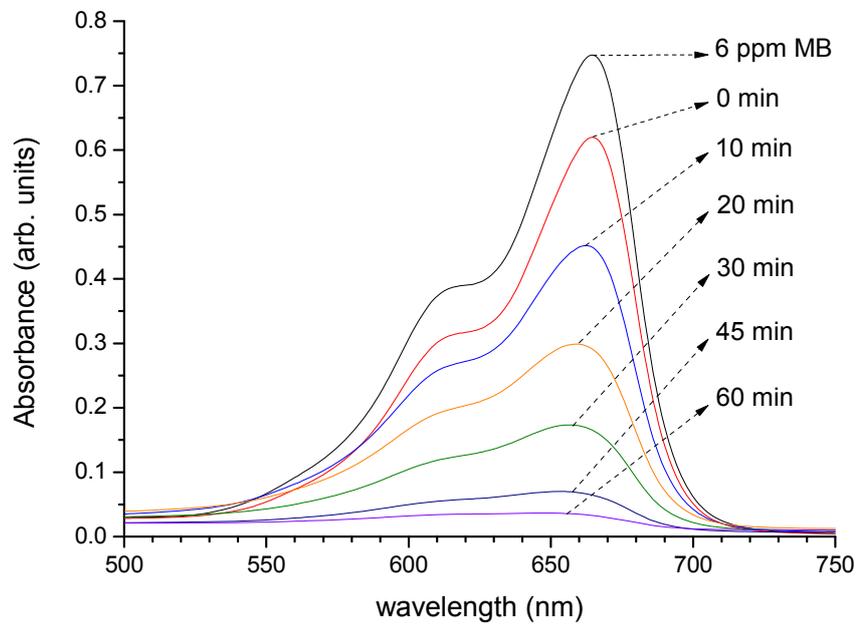

**Figure 9.** Absorption spectra of a 6 ppm MB solution during irradiation using ZnS/TNF as photocatalyst.

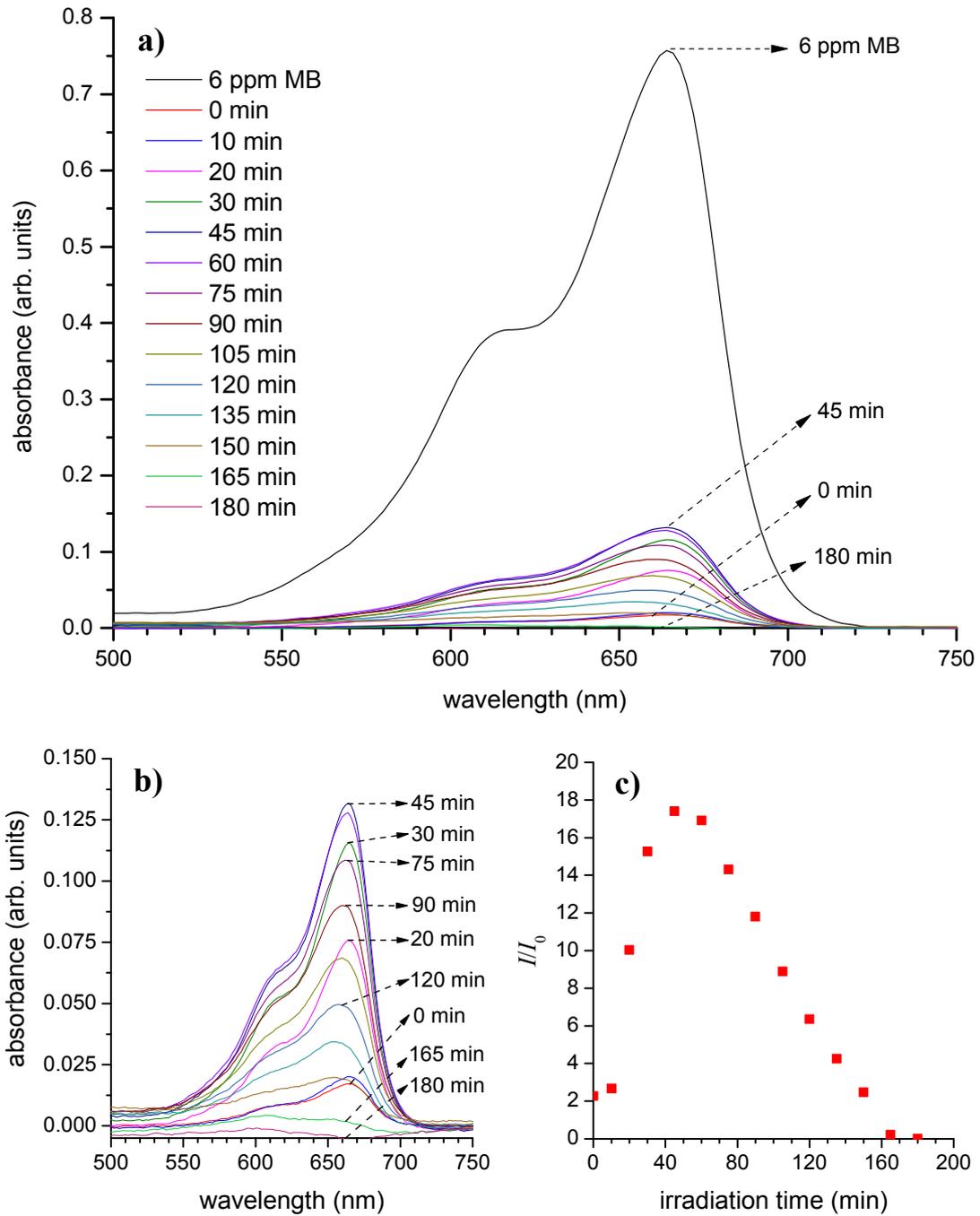

**Figure 10.** a) Absorption spectra of a 6 ppm MB solution before and during irradiation using ZnS(Bi$_2$S$_3$/TNF) as adsorbent and photocatalyst, b) detail of the spectra shown in a), and c) relative intensity of the MB chromophoric absorption band at λ = 664 nm as a function of the irradiation time. $I_0$ stands for the band intensity of the 6 ppm MB reference solution.